\documentclass[a4paper,11pt]{article}
\usepackage{jheppub}
\usepackage[T1]{fontenc}
\usepackage{amsfonts}
\usepackage{amsmath}
\usepackage{epsf}
\usepackage{amssymb}
\usepackage{latexsym}
\usepackage{graphicx,epsfig}
\usepackage{subfigure}
\usepackage[table]{xcolor}

\affiliation{Department of Physics, College of Science, Shiraz
University, Shiraz 71454, Iran \\ Biruni Observatory, College of
Science, Shiraz University, Shiraz 71454, Iran}
\emailAdd{amin.dehyadegari@hafez.shirazu.ac.ir}
\emailAdd{asheykhi@shirazu.ac.ir}
\abstract{We propose a modified gravity theory by extending the
Einstein-Hilbert action with an arbitrary function of the Ricci
scalar and the Kretschmann scalar invariants. The resulting
modified Friedmann equations for a spatially flat FRW universe are
derived, which remain free of higher-order derivatives and reduce
to the standard Friedmann equations in the limiting case.
Employing the gravity-thermodynamics conjecture, we investigate
the thermodynamic behavior at the apparent horizon and derive the
corresponding modified entropy. Using the first law of
thermodynamics together with the modified Friedmann equations, we
obtain a general expression for the apparent horizon entropy. This
formalism allows us to compute the modified entropy for various
well-known entropy models. Our approach establishes a consistent
thermodynamic framework linking modified gravity theories
constructed from curvature invariants to generalized entropy
functions on the cosmological apparent horizon.}

\begin{document}

\title{Modified Entropy from Action Principle}
\author{A. Dehyadegari, A. Sheykhi}
\maketitle

%%%%%%%%%%%%%%%%%%%%%%%%%%%%%%%%%%%%%%%%%%%%%%%%%%%%%%%%%%%%%%%%%%%%%%%%%%%%%%%%%%%%%%%%%%%%%%%%%%%%%%%%%%%%%%%%%%%%%%

\section{Introduction}

\label{Intro} The deep connection between gravity and thermodynamics has
been a subject of intense investigation ever since the pioneering works of
Bekenstein and Hawking, which revealed that black holes possess temperature
and entropy proportional to their horizon area. This insight was later
extended to cosmological settings, where it was shown that the Friedmann
equations governing the expansion of the universe can be derived from the
first law of thermodynamics applied to the apparent horizon \cite%
{CaiKim,Cai2,Shey1,Shey2}. Such a gravity-thermodynamics correspondence
suggests that gravitational field equations may be interpreted as a
thermodynamic equation of state, opening a new window to understand the
fundamental nature of gravity \cite{Jac,Elin,Pad1,Pad2,Pad3}.

In parallel, modified gravity theories have gained significant attention as
they offer possible explanations for the late-time accelerated expansion of
the universe without invoking dark energy, as well as for early-universe
inflation. Among these, curvature based modifications such as $f(R)$ gravity
have been extensively studied \cite{nojiri2011,capozziello2010,capozziello2011,nojiri2007}. However, less explored are modifications that
involve higher-order curvature invariants like the Kretschmann scalar $%
R_{\rho \sigma \mu \nu} R^{\rho \sigma \mu \nu}$ which naturally appear in
quantum gravity corrections and string theory effective actions.

In this work, we propose a novel modified gravity action for a spatially
flat Friedmann-Robertson-Walker (FRW) universe, where the Einstein-Hilbert
term is supplemented by an arbitrary function $f$ of a particular
combination of the Ricci scalar and the Kretschmann invariant, namely, $f=f(%
\mathcal{R}+\sqrt{6\mathcal{R}_{\rho \sigma \mu \nu }\mathcal{R} ^{\rho
\sigma \mu \nu }-\mathcal{R}^{2}})$. This combination is chosen for its
geometric simplicity and to avoid introducing higher-order derivatives in
the resulting field equations. The action contains an extension parameter $%
\lambda$, and when $\lambda=0$ standard general relativity is recovered.

We derive the modified Friedmann equations and then, following the
gravity- thermodynamics conjecture, study the thermodynamic behavior of the
apparent horizon. By employing the first law of thermodynamics along with
the continuity equation and the modified Friedmann equations, we obtain a
general expression for the entropy associated with the apparent horizon.
This entropy reduces to the standard area-law entropy in the limit $%
\lambda\rightarrow 0$ and can be explicitly computed for various
well-known entropy models by specifying the form of $f$. Recently,
deviations from the standard area-law entropy were obtained, using
a different approach based on stochastic fluctuations of the
spacetime metric \cite{khodahami2026}. Our results provide a
consistent thermodynamic interpretation of the proposed modified
gravity theory and may have implications for understanding the
microscopic origin of cosmological entropy.

The paper is organized as follows. In Sec. 2, we present the modified
gravity action and derive the corresponding modified Friedmann equations for
a flat FRW universe. In Sec. 3, we investigate the thermodynamic properties
of the apparent horizon, derive the modified entropy expression, and
summarize results for several entropy models in a table. Finally, Sec. 4
provides our conclusions and outlook.
%%%%%%%%%%%%%%%%%%%%%%%%%%%%%%%%%%%%%%%%%%%%%%%%%%%%%%%%%%%%%%%%%%%%%%%%%%%%%%%%%%%%%%%%%%%%%%%%%%%%%%%%%%%%%%%%%
\section{Modified gravity theory}

\label{sec:MGT} %\setcounter{equation}{0}
In this section, we propose a modified gravity action for the FRW universe
and derive the corresponding modified Friedmann equations. We begin by
considering the spatially flat line element of a homogeneous and isotropic
metric in $(3+1)$-dimensional spacetime, which is expressed as
\begin{equation}  \label{Eq:FRW}
ds^{2}=h_{ab}dx^{a}dx^{b}+R^{2}d\Omega ^{2},
\end{equation}
where $x^{0}=t$, $x^{1}=r$, $R=a(t)r$, and $a(t)$ is the scale factor. Here,
$h_{ab} = \text{diag}(-1, a^2(t))$ represents the metric of the
two-dimensional $(t, r)$ subspace, and $d\Omega ^{2}=d\theta
^{2}+\sin^2\theta d\phi ^{2}$ is the metric of a unit $2$-sphere.

Now, we generalize the standard Einstein-Hilbert action in $(3+1)$%
-dimensional spacetime with an arbitrary function $f$ as follows,
\begin{equation}
\mathcal{S}=\frac{1}{16\pi G}\int d^{4}x\sqrt{-g}\left[ \mathcal{R}+\lambda
f\!\left( \mathcal{R}+\sqrt{6\mathcal{R}_{\alpha \beta \mu \nu }\mathcal{R}%
^{\alpha \beta \mu \nu }-\mathcal{R}^{2}}\right) \right] +\mathcal{S}_{m},
\end{equation}
where $\mathcal{R}$ and $\mathcal{R}_{\alpha\beta\mu\nu}$ are the Ricci
scalar and Riemann curvature tensor, respectively. Additionally, $\lambda$
is an extension parameter and $\mathcal{S}_m$ stands for the matter action.
It is worth noting that the argument of the function $f$ is constructed from
a combination of the Ricci scalar and the Kretschmann scalar curvature
invariants. By varying the above action with respect to the metric tensor $%
g^{\mu\nu}$, we obtain the field equations in the following form:
\begin{equation}  \label{Eq:EoM}
G_{\mu \nu }+\lambda \,\mathcal{A}_{\mu \nu }=8\pi G\mathcal{T}_{\mu \nu },
\end{equation}
where $G_{\mu\nu}$ and $\mathcal{T}_{\mu\nu}$ are the Einstein and
stress-energy tensors, respectively. We do not show the expression of $%
\mathcal{A}_{\mu\nu}$ here, since it is rather lengthy and complicated. We
specify the matter content of the universe as a perfect fluid, whose
energy-momentum tensor is expressed as
\begin{equation}
\mathcal{T}_{\mu \nu }=\left( \rho +p\right) u_{\mu }u_{\nu }+pg_{\mu \nu },
\end{equation}
where $\rho$ and $p$ are, respectively, the energy density and pressure, and
$u_{\mu}=(1,0,0,0)$ is the four-velocity of the fluid. The matter
energy-momentum tensor is governed by the conservation law $\nabla _{\mu }%
\mathcal{T}^{\mu \nu }=0$, which leads directly to the continuity equation
\begin{equation}  \label{Eq:ConE}
\dot{\rho}+3H\left( \rho +p\right) =0,
\end{equation}
where the overdot symbol means the derivative with respect to the cosmic
time and $H = \dot{a}/a$ represents the Hubble parameter. Hence, by
substituting the flat FRW line element into the field equations %
\eqref{Eq:EoM}, the corresponding modified Friedmann equations take the
following form
\begin{eqnarray}\label{Eq:Frid}
&&H^{2}\left( 1-\frac{\lambda }{3}f^{\prime }\!\left( H^{2}\right) \right) +%
\frac{\lambda }{6}f\!\left( H^{2}\right) =\frac{8\pi G}{3}\rho, \\
&&\dot{H}\left( 1-\frac{\lambda }{6}f^{\prime }\!\left( H^{2}\right) -\frac{%
\lambda }{3}H^{2}f^{\prime \prime }\!\left( H^{2}\right) \right)
=-4\pi G\left( \rho +p\right),\label{Eq:Frid2}
\end{eqnarray}
where the prime is the derivative with respect to $H^{2}$. It should be
noted that the resulting modified equations remain free of higher-order
derivatives. Obviously, when $\lambda = 0$, the standard Friedmann equations
are recovered. In the next section, based on the gravity-thermodynamics
conjecture, we shall establish the thermodynamic laws at the apparent
horizon. By utilizing the conservation law along with the first modified
Friedmann equation, we will then derive the entropy associated with this
horizon.
%%%%%%%%%%%%%%%%%%%%%%%%%%%%%%%%%%%%%%%%%%%%%%%%%%%%%%%%%%%%%%%%%%%%%%%%%%%%%%%%%%%%%%%%%%%%%%%%%%%%%%%%%%%%%
\section{Modified entropy from action principle}

\label{sec:TMFEqs}\setcounter{equation}{0} In this section, based on the
deep connection between gravity and thermodynamics, we intend to examine the
thermodynamic behavior of the modified Friedmann equation and calculate the
corresponding thermodynamic quantities at the apparent horizon boundary. In
particular, we obtain the entropy expression associated with the apparent
horizon of the FRW universe.

In the cosmological context, the radius of the dynamical apparent horizon is
identified as the location where $h^{ab}\partial _{a}R_{h}\partial
_{b}R_{h}=0$. Consequently, the apparent horizon radius $R_{h}$ for a
spatially flat FRW universe is explicitly given by \cite%
{hayward1998,hayward1999,bak2000,akbar2006}
\begin{equation}  \label{Eq:AH}
R_{h}=1/H.
\end{equation}%
The associated temperature with the apparent horizon is determined by $%
T=\kappa /2\pi $, where $\kappa $ is the surface gravity on the apparent
horizon defined via \cite{akbar2006}
\begin{equation}
\kappa =\frac{1}{2\sqrt{-h}}\partial _{a}\left( \sqrt{-h}h^{ab}\partial
_{b}R_{h}\right) .
\end{equation}%
By using the FRW metric \eqref{Eq:FRW}, one obtains
\begin{equation}
T=-\frac{1}{2\pi R_{h}}\left( 1-\frac{\dot{R}_{h}}{2HR_{h}}\right) .
\label{Eq:Tem}
\end{equation}%
Following the gravity-thermodynamics correspondence, we write down the first
law of thermodynamics on the apparent horizon as
\begin{equation}
dE=TdS_{h}+WdV_{h},  \label{Eq:FLT}
\end{equation}%
where $E=\rho V_{h}$ is the total energy in volume $V_{h}=4\pi R_{h}^{3}/3$
enclosed by the apparent horizon, while $T$ and $S_{h}$ denote the
temperature and entropy of the apparent horizon, respectively. Here, the
quantity $W$ is the work density associated with the volume change of the
expanding universe given by \cite{akbar2006}
\begin{equation}
W=-\frac{1}{2}\mathcal{T}^{ab}h_{ab},
\end{equation}%
which in terms of the energy density and pressure reads as
\begin{equation}
W=\frac{1}{2}\left( \rho -p\right) .  \label{Eq:Work}
\end{equation}%
\begin{table}[t]
\centering
\rowcolors{3}{white}{gray!15}
\begin{tabular}{ccc}
\hline
&  &  \\[-2.3ex]
Entropy model & Modified Lagrangian & $S_{h}$ \\[0.7ex] \hline
&  &  \\[-2.3ex]
Kaniadakis \cite{nojiri2019,nojiri2022,sheykhi2024} & $\mathcal{R}-\lambda
X^{-1}$ & $S+\frac{\lambda G^2}{72\pi^2} S^3$ \\[0.8ex]
R\'enyi \cite{komatsu2017,moradpour2017} & $\mathcal{R}-\lambda \ln(X)$ & $S-%
\frac{\lambda G }{12\pi}S^2$ \\[0.8ex]
Logarithmic \cite{cai2008,xiao2022} & $\mathcal{R}-\lambda X^{2}$ & $S+\frac{%
    144\pi \lambda}{G} \ln(S)$ \\[0.8ex]
Inverse-area \cite{zhu2009corrections} & $\mathcal{R}+\lambda X^{3}$ & $S+%
\frac{4320\pi^2 \lambda}{G^2}S^{-1}$ \\[0.8ex]
Barrow \cite{barrow2020,sheykhi2021} & $\mathcal{R}-\lambda X(4-\ln(\frac{XG%
}{12\pi}))$ & $S+2\lambda S\ln(S)$ \\[0.8ex]
MOND \cite{sheykhi2025} & $\mathcal{R}-\lambda X^{1-n}$ & $S+\frac{%
2^{1-2n}(1-n)(1-2n)\lambda }{3^n(1+n)\pi^{n}G^{-n}} S^{1+n} $ \\%
[0.8ex] \hline
\end{tabular}%
\caption{The modified Lagrangian and the corresponding modified entropy $%
S_{h}$ for various well-known entropy models. The quantities are defined as $%
S=\protect\pi R_{h}^{2}/G$ and $X=\mathcal{R}+\protect\sqrt{6\mathcal{R}_{%
\protect\alpha \protect\beta \protect\mu \protect\nu }\mathcal{R}^{\protect%
\alpha \protect\beta \protect\mu \protect\nu }-\mathcal{R}^{2}}$.}
\label{tab:entropy}
\end{table}

We now proceed to derive the entropy of the apparent horizon based on the
modified Friedmann equations \eqref{Eq:Frid}. Taking the differential of the
total energy inside the apparent horizon, one obtains
\begin{align}
dE& =\dot{\rho}V_{h}dt+\rho dV_{h},  \notag \\
& =-4\pi R_{h}^{2}\left[ p+\rho (1-\dot{R}_{h})\right] dt,
\end{align}%
where the continuity equation \eqref{Eq:ConE} has been used in the second
equality. On the other hand, the first law of thermodynamics \eqref{Eq:FLT}
together with the expressions for the temperature \eqref{Eq:Tem} and the
work density \eqref{Eq:Work} yields
\begin{equation}
dE=-\frac{1}{2\pi R_{h}}\left( 1-\frac{\dot{R}_{h}}{2}\right) dS_{h}+2\pi
R_{h}^{2}\dot{R}_{h}\left( \rho -p\right) dt.
\end{equation}%
Combining the above expressions for the energy differential leads to
\begin{equation}
dS_{h}=8\pi ^{2}R_{h}^{3}\left( \rho +p\right) dt,
\end{equation}%
which, upon using the continuity equation \eqref{Eq:ConE}, reduces to
\begin{equation}
dS_{h}=-\frac{8}{3}\pi ^{2}R_{h}^{4}d\rho .  \label{Eq:Sh}
\end{equation}%
Next, rewriting the first Friedmann equation \eqref{Eq:Frid} in terms of the
apparent horizon radius $R_{h}$ using \eqref{Eq:AH} and taking its
differential yields
\begin{equation}
-2R_{h}^{-3}\left( 1-\frac{\lambda }{6}f^{\prime }\!\left( R_{h}^{-2}\right)
-\frac{\lambda }{3}R_{h}^{-2}f^{\prime \prime }\left( R_{h}^{-2}\right)
\right) dR_{h}=\frac{8\pi G}{3}d\rho ,
\end{equation}%
where the prime is the derivative with respect to $R_{h}^{-2}$. Finally,
substituting the above result into \eqref{Eq:Sh} and integrating, we find
the modified entropy of the apparent horizon as
\begin{equation}
S_{h}=S-\frac{\lambda G}{6\pi }\int S^{2}\left[ 3f^{\prime }(S)+2Sf^{\prime
\prime }(S)\right] dS,
\end{equation}%
where $S=\pi R_{h}^{2}/G$ is the standard entropy proportional to the
horizon area and the prime denotes differentiation with respect to $S$. Note
that the integration constant is set to zero to recover the standard
area-law entropy when $\lambda =0$. Thus, for a given functional form of $f$%
, the corresponding modified entropy of the apparent horizon can be
explicitly determined. The functional forms of $f$ corresponding to several
well-known entropy models are summarized in Table \ref{tab:entropy}.

%%%%%%%%%%%%%%%%%%%%%%%%%%%%%%%%%%%%%%%%%%%%%%%%%%%%%%%%%%%%%%%%%%%%%%%%%%%%%%%%%%%%%%%%%%%%%%%%%%%%%%%

\section{Conclusions and outlook}

In this work, we have proposed a modified gravity theory in a spatially flat
FRW universe by extending the standard Einstein-Hilbert action with an
arbitrary function of the Ricci scalar and the Kretschmann scalar invariant,
specifically $f=f(\mathcal{R}+\sqrt{6\mathcal{R}_{\rho \sigma \mu \nu }%
\mathcal{R}^{\rho \sigma \mu \nu }-\mathcal{R}^{2}})$. The inclusion of this
particular combination ensures that the resulting modified Friedmann
equations remain free of higher-order derivatives, a desirable feature for
maintaining a well-defined initial value problem. The modified field
equations contain an extension parameter $\lambda$, and in the limit $%
\lambda\rightarrow0$, the standard Friedmann equations of general relativity
are recovered.

Using the gravity-thermodynamics conjecture, we examined the thermodynamic
behavior of the apparent horizon of the FRW universe. By applying the first
law of thermodynamics $dE=TdS+WdV$ together with the continuity equation and
the modified Friedmann equations, we derived a general expression for the
entropy associated with the apparent horizon
\begin{equation}
S_{h}=S-\frac{\lambda G}{6\pi}\int S^{2}\left[ 3f^{\prime }(S)+2Sf^{\prime
\prime }(S)\right] dS,
\end{equation}
where $S=\pi R_{h}^2/G$ is the standard Bekenstein-Hawking entropy
associated with the apparent horizon. This expression explicitly shows how
the modified gravity action leads to corrections to the area-law entropy.
For a given functional form of $f$, the corresponding modified entropy can
be computed. The results for several well-known entropy models
(e.g.,R\'enyi, Barrow, Kaniadakis, etc.) are summarized in Table 1,
demonstrating the versatility of our approach.

Several interesting directions remain for future investigation. First, it
would be valuable to explore the cosmological implications of the modified
Friedmann equations derived here, such as the possibility of inflation or
late-time acceleration without exotic matter components. Second, the
explicit form of the tensor $\mathcal{A}_{\mu \nu}$ in the field equations,
though lengthy, could be analyzed for specific symmetric spacetimes to
reveal additional physical effects. Third, extending this framework to
include spatial curvature or to non-flat FRW universes would provide a more
complete thermodynamic picture. Fourth, the connection between the modified
entropy expression and quantum gravity corrections-such as those arising
from loop quantum gravity or string theory-deserves further study. Finally,
observational constraints on the parameter $\lambda$ and the functional form
of $f$ could be derived from cosmological data, offering potential tests of
the model. Our work thus provides a foundation for a deeper understanding of
the interplay between modified gravity and the thermodynamics of spacetime.
%%%%%%%%%%%%%%%%%%%%%%%%%%%%%%%%%%%%%%%%%%%%%%%%%%%%%%%%%%%%%%%%%%%%%%%%%%%%%%%%%%%%%%%%%
\acknowledgments{We thank Shiraz University Research Council. This
work is based upon research funded by Iran National Science
Foundation (INSF) under project No. 4028419.}
%%%%%%%%%%%%%%%%%%%%%%%%%%%%%%%%%%%%%%%%%%%%%%%%%%%%%%%%%%%%%%%%%%%%%%%%%%%%%%%%%%

\end{document}